\documentclass[12pt,a4paper,twoside]{article}
\usepackage{epsfig}
\usepackage{amssymb,latexsym,amsfonts}

\newcommand{\be}{\begin{equation}}
\newcommand{\ee}{\end{equation}}
\newcommand{\bd}{\begin{displaymath}}
\newcommand{\ed}{\end{displaymath}}
\newcommand{\ba}{\begin{array}}
\newcommand{\ea}{\end{array}}
\newcommand{\bt}{\begin{tabular}}
\newcommand{\et}{\end{tabular}}

\newcommand{\bea}{\begin{eqnarray}}
\newcommand{\eea}{\end{eqnarray}}
\newcommand{\bean}{\begin{eqnarray*}}
\newcommand{\eean}{\end{eqnarray*}}

\newcommand{\hlf}{\frac{1}{2}}

\newcommand{\cV}{{\cal V}}

\newcommand{\cH}{{\cal H}}

\newcommand{\dif}{\mathrm{d}}

\newcommand{\inp}[2]{\langle #1, #2 \rangle}

\newcommand{\Z}{\mathbb{Z}}

\newcommand{\R}{\mathbb{R}}
\newcommand{\C}{\mathbb{C}}
\newcommand{\h}{\mathbb{H}}

\newcommand{\ov}[1]{\overline{#1}}
\newcommand{\df}[1]{{} * \! #1}

\begin{document}
\begin{titlepage}

\begin{center} 
\today \hfill                  hep-th/0212024

\hfill VUB-TENA-02-07

\vskip 2 cm
{\Large \bf The group theory of oxidation II:\\
\vskip.5cm
Cosets of non-split groups}\\
\vskip 1.25 cm
{Arjan Keurentjes\footnote{email address: Arjan@tena4.vub.ac.be}}\\
\vskip 0.5cm
{\sl Theoretische Natuurkunde, Vrije Universiteit Brussel,\\ Pleinlaan
  2, B-1050 Brussels, Belgium \\}
\end{center}
\vskip 2 cm
\begin{abstract}
\baselineskip=18pt
The oxidation program of reference \cite{Keurentjes:2002xc} is extended to
cover oxidation of 3-d sigma model theories on a coset $G/H$, with $G$
non-compact (but not necessarily split), and $H$ the maximal compact
subgroup. We recover the matter content, the equations of motion and
Bianchi identities from group lattice and Cartan involution. Satake
diagrams provide an elegant tool for the computations, the maximal
oxidation dimension, and group disintegration chains can be directly
read off. We  give a complete list of theories that can be recovered
from oxidation of a 3 dimensional coset sigma model on $G/H$, where $G$
is a simple non-compact group. 
\end{abstract}

\end{titlepage}
\section{Introduction}

Upon dimensional reduction of a physical theory containing gravity to
3 dimensions, and after appropriate dualizations, all degrees of
freedom can be represented by scalars. The opposite of dimensional
reduction is ``oxidation'': Interpreting a theory as a reduction of a
higher dimensional one, and reconstructing this higher dimensional
theory from the lower dimensional one. In this paper we will be
interested in 3 dimensional sigma models on $G/H$, where $G$ is a
non-compact group and $H$ its maximal compact subgroup, and in the
theories that can be oxidized from them.

In \cite{Breitenlohner:1987dg} a list of 3 dimensional theories based
on coset spaces that are oxidizable appeared, together with their 4
dimensional counterparts. In \cite{Cremmer:1999du} an analysis of
oxidation from 3 dimensional theories on coset spaces of the form
$G/H$, with $G$ split (maximally non-compact) was presented; many such
theories can actually be oxidized to more than 4 dimensions. An
interesting aspect of oxidation is that it is not unique; there may be
different higher dimensional theories leading to the same lower
dimensional theory. In \cite{Henry-Labordere:2002dk} and
\cite{Keurentjes:2002xc} such branches are analyzed from
different points of view. The paper \cite{Henry-Labordere:2002dk}
builds on developments from \cite{Cremmer:1998px} and explores a
suggestive link with Del Pezzo surfaces \cite{Iqbal:2001ye}. In
\cite{Keurentjes:2002xc} we connected some old ideas
\cite{Julia:1980gr, Julia:1982gx} to a systematic recipe for
dimensional reduction as developed in \cite{Lu:1995yn, Cremmer:1997ct,
  Cremmer:1999du}. Via a recasting of the sigma model equations, we were
able to do an exhaustive analysis of theories with split groups, using
mainly Lie group theory, and we rederived (and extended) the results of
\cite{Cremmer:1999du} from this unified perspective. 

It seems that so far there was no systematic understanding for
theories oxidized from cosets that are not based on split groups. It
is hard to deal with these with the methods of
\cite{Henry-Labordere:2002dk}, as it is not (yet) clear how to connect
the Del Pezzo/gravity correspondence to groups that are not subgroups
of $E_{8(8)}$. The methods of \cite{Keurentjes:2002xc} however can be
straightforwardly extended to this more general case, and this will be
the topic of this paper. 

Most of the theories we find appear to have been described already in
the literature. Our analysis is systematic, and includes for example
the bosonic sectors of the $N=2$ theories related by the $\mathbf{r}$-
and $\mathbf{c}$-maps \cite{Cecotti:1988qn, Cecotti:1988ad, deWit:1991nm,
  deWit:1992wf}, insofar as these give rise to symmetric spaces (the
other spaces result in higher dimensional theories with supergravity
as a subsector, and therefore must have subspaces that are symmetric spaces).

In section \ref{oxi} we will revisit and extend the methods of
\cite{Keurentjes:2002xc} to cover cosets from generic non-compact
groups. Apart from the introduction of some extra elements from group
theory, the analysis is very similar to the one for the split
groups. Many of the results from \cite{Keurentjes:2002xc} carry over
straightforwardly. 

In section \ref{Satake} we discuss the diagrammatic language of
Satake diagrams. These are helpful with the analysis of the
cosets as they encode all the relevant information about the
non-compact group. They also capture the full process of ``group
disintegration''; in particular the maximal oxidation dimension, and
the non-compact forms appearing in the oxidation procedure can be
directly read off. 

In section \ref{results} we analyze theories oxidized from non-compact
cosets, including $G/H$ with $G$ a group over the complex
numbers, and all cosets based on non-compact simple groups.

In a final section we summarize, and highlight some significant points.

Almost all our conventions regarding Lie groups can be found in the
appendices of \cite{Keurentjes:2002xc}. Necessary additional material will be
introduced in section \ref{oxi}.

\section{General theory of oxidation}\label{oxi}

Our philosophy is mainly based on ideas from \cite{Julia:1980gr} (where
they were discussed in the context of supergravities) and can be found
in section 2 of \cite{Keurentjes:2002xc}. These ideas are
quite general, and will also be the basis for our analysis here. We
briefly repeat the main points, and refer to section 2
of \cite{Keurentjes:2002xc}, and references therein for more details.

The starting point is a coset sigma model in 3 dimensions on $G/H$,
with $G$ a non-compact group, and $H$ its maximal compact
subgroup (one can include the pathological case where $G$ is
compact; then $G=H$, the coset is just a point and the sigma model is
empty). The dimensional reduction of general relativity from
$D$-dimensions to 3 dimensions results in a sigma model on
$SL(D \! - \! 2,\R)/SO(D \! - \! 2)$ coset
\cite{Cremmer:1997ct}. Interpreting the $G/H$-model in 3 dimensions as
resulting from some higher dimensional theory, the gravity sector can
be recovered by studying the embeddings of $SL(D \! - \! 2,\R) \subset
G$. Decomposing $G$ into $SL(D \! - \! 2,\R)$ irreducible
representations (irreps), we find at least the adjoint irrep of $SL(D
\! - \! 2,\R)$, to be interpreted as the graviton. We demand exactly
one graviton, and want to interpret the remaining irreps as massless
fields (forms). This translates into a constraint on the possible
embeddings of $SL(D \! - \! 2,\R)$ in $G$: They have to be \emph{index 1}
embeddings \cite{Keurentjes:2002xc}. Equivalently, the roots of
$SL(D \! - \! 2,\R)$ can be chosen such that they coincide with \emph{long} roots
of $G$.

As the adjoint irrep of $G$ is a self-conjugate representation, we
must find self conjugate representations of $SL(D \! - \! 2,\R)$ in the
composition. These come in two kinds: Self-conjugate irreps, and pairs
of mutually conjugate irreps. The self-conjugate irreps are, due to
the restriction to level 1 subgroups, the adjoint (representing the graviton),
singlets (scalars), and self-conjugate $(D-2)/2$ tensors (if $D$ is
even; we will discuss these momentarily). The pairs
of mutually conjugate irreps, are always combination of an $n$-form
and a $(D-2-n)$ form; these represent a form field and its dual. In
\cite{Keurentjes:2002xc}, we associated an equation to every
$SL(D \! - \! 2,\R)$-irrep: the adjoint irrep is linked to the
Einstein equation, every $n$-form gives a Bianchi identity for a form
field $F_{(n+1)}$, the $(D-2-n)$-form gives the equation of
motion for the same form. There is no fundamental distinction between
the $n$- and the $(D-2-n)$-form; this is how the possibility to
dualize fields is built in in our theory. A self-conjugate tensor
irrep gives rise to either a self-dual tensor for $D-2=4k$, or
``pseudo-selfduality'' (with the imaginary unit $i$ occurring in the
duality equation) for $D=4k$; in both cases one has only half the
number of equations, and degrees of freedom. 

The centralizer of $SL(D \! - \! 2,\R)$ in $G$ acts as a symmetry group
on the theory. We call this group the ``U-duality group'', and denote it
by $U_D$. 

Recovering the level 1 embeddings of $SL(D \! - \! 2,\R)$ in $G$ is quite easy
if $G$ is split. Then one can pick a basis for the Lie algebra of $G$,
consisting of Cartan generators $H_i$, and ladder operators $E_{\pm
  \alpha}$, such that $G$ is generated by linear combinations of these
generators with \emph{real} coefficients. Picking a sublattice of $G$,
the corresponding generators generate a group that is also split;
picking a $A_{D-3}$ sublattice of long roots, the corresponding
generators generate $SL(D \! - \! 2,\R)$ (and not another real
form). This last statement is not true for generic non-compact groups,
and requires us to introduce some more technology.  

\subsection{Non-compact groups}

A notion that is central to the study of non-compact real forms of
 semi-simple Lie-groups is that of a Cartan involution
 \cite{Helgason}. From the Cartan involution the non-compact real form
 of the group can be  easily reconstructed.

An involutive automorphism $\theta$ is called a Cartan
involution if $-\inp{X}{\theta Y}$ is strictly positive definite for
all algebra generators $X,Y$. An involution has eigenvalues $\pm 1$,
and the realization of the involution can be chosen such that the
Cartan subalgebra is closed under the involution. 

Under the action of $\theta$ on the Cartan subalgebra generators
$\alpha = \alpha_i H_i$, the root space $\cH$ decomposes into two
orthogonal complements. The space spanned by eigenvectors of $\theta$
with eigenvalue 1 we call $\cH_+$, its complement $\cH_-$. The lattice 
in the subspace $\cH^+$ containing the invariant roots
$\theta(\alpha)=\alpha$ is the lattice of a compact group $G_c$. We
denote  the set of invariant roots by $\Delta_c$. Similarly, we use
$\Delta^+_c$ for $\Delta^+ \cap \Delta_c$. The generators of $G_c$ are
$iH_j$, $E_{\alpha}-E_{-\alpha}$, and $i(E_{\alpha}+E_{-\alpha})$. We
emphasize however that $G_c$ is \emph{not} a maximal compact subgroup;
it is not even a maximal regular compact subgroup. It will however
play a crucial role in our construction of the sigma model. 

The regular subgroup commuting with $G_c$ will be called $G_s$. The
roots of $G_s$ obey $\theta(\alpha)=-\alpha$, and the set of roots of
$G_s$ will be denoted by $\Delta_s$. The generators of $G_s$ are of
the form $H_j$, $E_{\alpha}-E_{-\alpha}$, and
$E_{\alpha}+E_{-\alpha}$. The group $G_s$ is a regular split group; it
need however not be the maximal regular split group (a counterexample
is found for $SO(p,q)$ with $p$ or $q$ odd) nor does it imply that it
is an index 1 subgroup (a counterexample is found for
$F_{4(-20)}$). We recover the index 1 $SL(D \! - \! 2, \R)$-subgroups
required for oxidation as subgroups of $G_s$; nevertheless, the group
$G_s$ is not as important to our analysis as $G_c$. 

The remaining roots mix under the Cartan involution. We define the
image of the ladder operator $E_{\alpha}$ to be $C_{\alpha}
E_{\theta(\alpha)}$, with $C_{\alpha} = \pm 1$, where the plus- or
  minus sign should be chosen consistently with a number of conditions,
  that we will describe now. First of all, if $\alpha \in \Delta_c$,
  $C_{\alpha}=1$. If $\alpha \in \Delta_s$, we choose
  $C_{\alpha}=-1$. If $\alpha \notin (\Delta_c + \Delta_s)$, closure
  of the algebra combined with the fact that $\theta$ is an automorphism
  leads to the requirement
$$
C_{\alpha+\beta} N_{\alpha,\beta} = C_{\alpha} C_{\beta}
N_{\theta(\alpha),\theta(\beta)}
$$
This may still leave some signs unfixed, but this is not important for
what follows.

To specify a non-compact real form, one starts from the compact real
form, and a Cartan involution. With the above considerations, it
suffices to specify the action on the root space 
(and fix a convention for the $C_{\alpha}$). We then divide the
generators of the compact form in a set invariant under the Cartan
involution, and a set that has eigenvalue $-1$. As generators for the
corresponding real form, we take the invariant generators, and add to
these the non-invariant generators multiplied by the imaginary unit
$i$. The invariant generators generate a compact subgroup; the
non-invariant generators turn into ``non-compact'' generators by
multiplication by $i$. Notice that in the extreme cases that $G$ is
compact, or $G$ is split, one immediately recovers the generators
specified previously.

Two important characteristics of non-compact groups are the $\R$--rank
(real rank) and the character.

The $\R$--rank can be defined as follows: A Cartan
  subalgebra\footnote{We use the same symbol for the root space and
  the Cartan subalgebra, as the two can be identified.} $\cH$ is
  a maximal Abelian subalgebra with $\textrm{ad}(\cH)$ completely
  reducible. The Cartan subalgebra generates an Abelian group called a
  \emph{torus} (Beware: in this context a torus has the topology
  $(S^1)^m \times \R^n$ for some $m,n \geq 0$, and hence is only a
  torus in the usual sense if $n=0$). A torus is $\R$-split if it is
  diagonalizable over $\R$ (and hence one which has $m=0$ instead of
  $n$!). The $\R$-rank is defined as the dimension of a maximal
  $\R$-split torus. It equals the multiplicity of the eigenvalue $-1$
  of the restriction of $\theta$ to ${\cal H}$, and therefore
  $\textrm{dim}(\cH_-)$. The $\R$-rank is maximal for the split form
  of a group, and then coincides with the rank. Its minimal value is
  zero, for compact forms. 

The character is more easily defined. Let $G$ be a non-compact group,
and $H$ its maximal compact subgroup. Then there are $d_-
=\textrm{dim}(H)$ compact generators, while the complement consists of
$d_+ = \textrm{dim}(G)-\textrm{dim}(H)$ ``non-compact
generators''. The character $\sigma$ is defined as $\sigma =
d_+ - d_-$. The maximal value of the character coincides with the
rank, and is obtained for the split groups (and then automatically
equals the $\R$-rank). The minimum value for the character is obtained
if $d_+ =0$, hence for compact $G$, when $\sigma = - \textrm{dim}(G)$.

Suppose we start with a coset sigma model on $G/H$ in 3 dimensions,
with $G$ a group with character $\sigma_3$ and $\R$-rank $r_3$. If
this theory is oxidizable to $D$ dimensions, one finds a U-duality
group with character $\sigma_D$ and $\R$-rank $r_D$ are given
by \cite{Julia:1980gr} 
\be
\sigma_{D} = \sigma_3 - (D-3); \qquad r_D= r_3 - (D-3).
\ee 
The statement on the character takes into account the possible
presence of compact factors. Such factors are not manifest in the
sigma model (they drop out after division by the maximal compact
subgroup), but do impose important restrictions on the matter sectors
(that have to organize in representations of these compact
factors). Omitting such compact factors leads to a break-up of the
pattern. With this proviso, the above statements are actually easy to
prove.

Consider the algebra elements $H_i$, $E_{\pm \alpha}$, and the Cartan
involution acting on the root space. The oxidation recipe demands a
level 1 $A_{D-3}$ sublattice of the root lattice; because $SL(D \! - \! 2)$ is
split, this lattice is contained in a subspace on which the Cartan
involution $\theta$ acts as $- \mathbf{1}$. Hence on the complementary
subspace, the Cartan involution acts
$\textrm{diag}(1,\ldots,1,-1,\ldots,-1)$, where the number of minus
signs is $r_D = r_3 -(D-3)$. 

Regarding the character of the groups, let us decompose the root
lattice. The root lattice of $SL(D \! - \! 2)$ is a sublattice, the
complementary subspace has the lattice of $U_D$, and there are a
number of roots that have components in both subspaces. Let us start with
the latter. Consider a root $\alpha$ of $G$, where $\alpha$ has
components in the direction of the $SL(D \! - \! 2)$ lattice. The Cartan
involution $\theta$ acts as $-\mathbf{1}$ on the subspace containing the
$SL(D \! - \! 2)$, and we can form the generators $E_{\alpha} \pm C_{\alpha}
E_{\theta(\alpha)}$. The combination with the plus sign is invariant
  under the Cartan involution, and hence corresponds to a compact
  generator (upon combination with its hermitian
  conjugate), while the combination with the minus sign gives a
  non-compact generator. Important is that in this way, we find that
  roots with  components in both subspaces give rise to equal numbers
  of compact and non-compact generators, and hence do not contribute
  to the character. Hence:  
\be
\sigma_3 = \sigma(G) = \sigma(U_D) + \sigma(SL(D \! - \! 2)) = \sigma_D+(D-3).
\ee     
Note that regularity of the subalgebra's plays an important role in the proof.

\subsection{Sigma models on non-compact cosets}

As in \cite{Keurentjes:2002xc}, we base our discussion on the
following form of the sigma model action ($\cV \in G$)
\bea
{L}_{G/H} = -e \ \textrm{tr}\left((\partial \cV)  \cV^{-1}\hlf(1+T)(\partial
\cV) \cV^{-1}\right). \label{taction}
\eea
The operator $T$ acts on algebra elements $A$ as $T(A)=-\theta(A)$;
because $\theta$ is an involution $\hlf(1+T)$ is a projection
operator.

The form $(\dif \cV)\cV^{-1}$ can be expanded in generators as follows:
\be \label{tform}
(\dif \cV) \cV^{-1} = \hlf \sum_{i=1}^r \dif {\phi}^i {H}^i +
\sum_{\alpha \in  \Delta^+_{nc}} e^{\hlf{\inp{\alpha}{\phi}}}
F_{(1)\alpha} E_{\alpha}.  
\ee 
This expression assumes a particular gauge, that is implicit in the
choice of symbols. The constant $r$ denotes the $\R$-rank; the $H_i$
form a basis for a maximal $\R$ split torus. The symbol
$\Delta^+_{nc}$ denotes $\Delta^+ - \Delta^+_c$, the set of
positive roots of $G$ that are not roots of $G_c$. That one can choose
this gauge follows from the Iwasawa decomposition\footnote{This way of
  parameterizing the sigma model leads to the identity  
$$ 
\dim(G) - \dim(H) = \hlf(\dim(G) + r(G)) - \hlf (\dim(G_c) + r(G_c)).
$$
The ranks of $G,G_c$ are denoted by $r(G), r(G_c)$. Left and right
hand side express the number of scalars, the left hand side from the
abstract definition, the right hand side from the gauge
choice; the two terms in brackets give the number of positive roots
plus the dimension of the Cartan subalgebra of $G$, and $G_c$.}. Equation
(\ref{tform}) is almost identical to equation (25) in
\cite{Keurentjes:2002xc}; for split $G$ it is identical, as then $r$
coincides with the rank, and $\Delta^+_c = \emptyset$.    

Another modification is that $\phi$ only has components for the
directions corresponding to the $\R$-split torus. Because there are
fewer dilatons, the inner products $\inp{\alpha}{\phi}$ are
defined by setting those components of $\phi$ in the
direction transverse to the $\R$-split torus to zero. An amusing
observation is that the dilatons can be identified with elements of
the Cartan subalgebra; elements of the Cartan subalgebra with
negative norm (compact generators) would lead to kinetic terms with
wrong signs, but are projected out by the denominator compact subgroup.

All formula's of the sections 4.1 and 4.2 of \cite{Keurentjes:2002xc}
can be copied to the more general case, by setting $r$ to be the
$\R$-rank, replacing $\Delta^+$ by $\Delta^+_{nc}$, and remembering to
modify the inner products involving $\phi$ by setting the appropriate
components of $\phi$ to zero. These modifications do not change any
derivation from \cite{Keurentjes:2002xc}. 

Hence we find the Bianchi identities:
\be \label{scbi}
\dif F_{(1)\gamma} = \hlf \sum_{*} N_{\alpha,\beta} F_{(1)\alpha}
\wedge F_{(1)\beta} \qquad * = \left\{\ba{l} \alpha, \beta, \gamma \in
\Delta^+_{nc}\\  
\alpha+ \beta = \gamma \ea \right.;
\ee
and the equations of motion:
\be \label{sceqmo}
\dif F_{(D-1)-\gamma}= \sum_*  N_{\alpha,-\beta} F_{(1)\alpha} \wedge
F_{(D-1)-\beta}\qquad * =\left\{\ba{l} \alpha,\beta,\gamma \in
\Delta^+_{nc}\\  \alpha-\beta=-\gamma \\
\ea \right. ,
\ee
where we have defined
\be \label{defmin}
F_{(D-1)-\gamma} \equiv e^{\inp{\gamma}{\phi}}\df{F}_{(1)\gamma} =
\dif A_{(D-2)-\gamma}-\sum_{\beta-\alpha=-\gamma} N_{\beta,-\alpha}
F_{(1)\beta}  \wedge A_{(D-2)-\alpha} .
\ee

It seems there is no longer a one on one relation between algebra
generators and equations, as in \cite{Keurentjes:2002xc}. This
relationship is recovered when supplementing the equations of motion
and Bianchi identities with algebraic equations for the missing
generators:
\be \label{alg}
\phi_i =0 \ \textrm{for }i > r \qquad F_{(n)\alpha} = 0 \ \textrm{for
}\alpha \in \Delta_c.
\ee
These equations are a consequence of the Iwasawa decomposition,
reflected in the  choice of gauge. For $G$ compact, all equations are
algebraic, and the model empty. 

Also the addition of matter proceeds analogously to
\cite{Keurentjes:2002xc}. The matter Bianchi identity becomes
\be \label{bi2}
\dif F_{(n)\lambda'}= \sum_{*} N_{\alpha,\lambda} F_{(1)\alpha} \wedge
F_{(n)\lambda} \qquad * = \left\{\ba{l} \lambda, \lambda' \in
\Lambda\\ 
\alpha \in \Delta^+_{nc}\\
\alpha+ \lambda = \lambda' \\
\ea \right.;
\ee
and the equation of motion becomes
\be \label{eqmo2}
\dif F_{(D-n)-\lambda'}= \sum_{*} N_{\alpha,-\lambda} F_{(1)\alpha}
\wedge F_{(D-n)-\lambda} \qquad * = \left\{\ba{l} \lambda,\lambda' \in
\Lambda \\ 
\alpha \in \Delta^+_{nc}\\
\alpha-\lambda=- \lambda' \\
\ea \right. .
\ee

Again the set of weights $\Lambda$ belongs to a certain $G$
representation, and $\ov{\Lambda}$ to the conjugate representation.
We remind the reader of the possibility that form and dual form
transform in a self-conjugate representation, which must be realized
in theories with self-dual forms. Then the
equation of motion (\ref{eqmo2}) and Bianchi identity (\ref{bi2}) are
the same equation, and we can consistently impose self-duality. 

Again we have one Bianchi identity, and one equation of
motion, labelled by $\lambda$ respectively $-\lambda$. For self-dual
representations constraint equation and Bianchi identity imply each
other, but as $\lambda$ and $-\lambda$ belong to the same
representation we precisely get as many equations as weights. 

The equation of motion for the dilatonic scalars becomes
\be \label{cartaneqmo2}
2 \dif(\df{\dif \phi}^i) = \sum_{\alpha \in \Delta^+_{nc}} \alpha^i
F_{(D-1)-\alpha} \wedge F_{(1)\alpha} + \sum_{\lambda \in \Lambda}
\lambda^i  F_{(D-n)-\lambda} \wedge F_{(n)\lambda} . 
\ee
In absence of matter, the second sum drops out.

\subsection{Oxidation}

The equations of the previous section are almost identical to the ones
appearing in \cite{Keurentjes:2002xc}. It should come as no surprise
that also the oxidation recipe is hardly modified.

To be precise, there is only one small modification in the oxidation
recipe for the axions. The assignement of forms to the antisymmetric
tensor representations of $SL(D-2)$ proceeds as in
\cite{Keurentjes:2002xc}. The difference is in the singlets of
$SL(D \! - \! 2)$; these correspond to the the group $U_D$. In
\cite{Keurentjes:2002xc} we made a positive root decomposition of the
semi-simple part of $U_D$, while choosing positive directions for the
Abelian factors. Here we do the same, but on top of that we have to
identify the roots of the group $G_c \subset U_D$. We call
this set of roots $\Delta_c$. We again associate 1-forms
$F_{(1)\alpha}$ to positive roots, and $D-1$-forms $F_{(D-1)-\alpha}$
to the negative roots. If $\alpha$ is a root of $\Delta_c$, then so is
$-\alpha$, and we set 
\be
F_{(k)\pm\alpha}=0, \qquad k=1,D-1.
\ee

The remaining equations are again given by
\be \label{oxieq}
\dif F_{(n)\alpha'} = \hlf \sum_* \eta_{l,\beta;m,\gamma} \
N_{\beta,\gamma}  F_{(l)\beta'} \wedge F_{(m)\gamma'}  \qquad * =
\left\{\ba{l} l+m=n+1\\  \alpha'+ \beta' = \gamma'
\ea \right. .
\ee
Computation of the sign factors $\eta_{l,\beta;m,\gamma}$ proceeds
as explained in \cite{Keurentjes:2002xc}. We repeat once more that the
only difference is in exponential prefactors not manifest in
(\ref{oxieq}); we have
\be
F_{(D-n)-\alpha} \equiv e^{\inp{\alpha}{\phi}} \df{F}_{(n)\alpha}
\ee
with $\phi$ including some components set to zero from
the start. 

The dilaton equation is always (\ref{cartaneqmo2}), where $i$ is
restricted to run from 1 to the $\R$-rank $r$: Components transverse
to the $\R$-split torus play no role. 

It should be clear that our comparison of this oxidation recipe to
dimensional reduction, in section 4.5 of \cite{Keurentjes:2002xc}
requires no repetition or modification.

\section{Satake diagrams} \label{Satake}

So far we have used the abstract description of non-compact groups and
their algebra's. For specific computations, we have to specify the
(complexified) algebra, and the Cartan involution. These
can be encoded in so-called Satake diagrams \cite{Helgason}.

\subsection{Non-compact groups from diagrams}

\begin{figure}[ht]
\begin{center}
\includegraphics[bb=0 600 500 750,width=8cm]{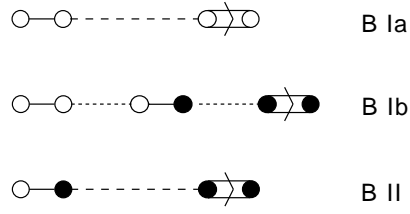}
\caption{Satake diagrams for non-compact forms of the $B_n$
  groups}\label{SB} 
\end{center}
\end{figure}

\begin{figure}[ht]
\begin{center}
\includegraphics[bb=0 600 500 750,width=8cm]{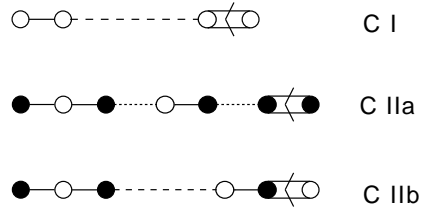}
\caption{Satake diagrams for non-compact  forms of the $C_n$
  groups}\label{SC} 
\end{center}
\end{figure}

\begin{figure}[ht]
\begin{center}
\includegraphics[bb=0 270 500 750, width=8cm]{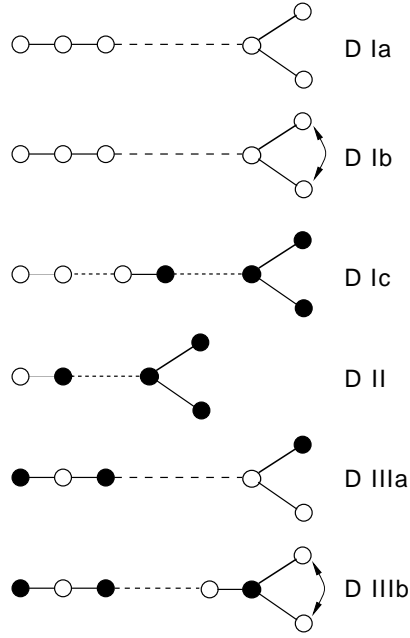}
\caption{Satake diagrams for non-compact forms of the $D_n$
  groups}\label{SD} 
\end{center}
\end{figure}

\begin{figure}[ht]
\begin{center}
\includegraphics[bb=0 480 500 800, width=8cm]{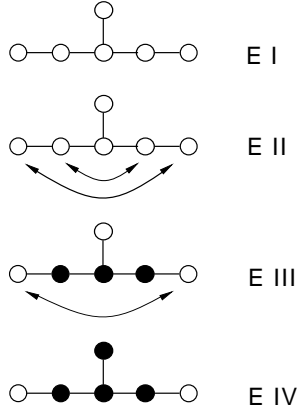}
\caption{Satake diagrams for non-compact forms of the $E_6$}\label{SE6}
\end{center}
\end{figure}

\begin{figure}[ht]
\begin{center}
\includegraphics[bb=0 620 500 800,width=8cm]{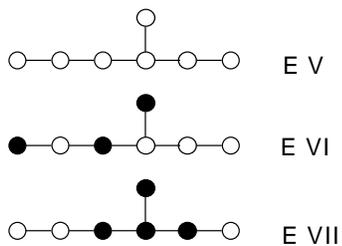}
\caption{Satake diagrams for non-compact forms of the $E_7$
  groups}\label{SE7} 
\end{center}
\end{figure}

\begin{figure}[ht]
\begin{center}
\includegraphics[bb=0 700 500 750, width=8cm]{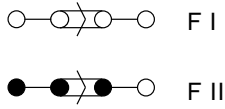}
\caption{Satake diagrams for non-compact forms of $F_4$}\label{SF}
\end{center}
\end{figure}

\begin{figure}[ht]
\begin{center}
\includegraphics[bb=0 745 500 755, width=8cm]{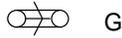}
\caption{Satake diagrams for non-compact forms of $G_2$} \label{SG}
\end{center}
\end{figure}

A Satake diagram is a Dynkin diagram, with additional
``decoration'' (Satake diagrams for all simple real forms can be found
in the figures \ref{SA}-\ref{SG}). In a traditional Dynkin diagram
all nodes are of the same colour, in a Satake diagram we use black
(solid) dots for some nodes, and white (open) dots for others. On top
  of that, there is the possibility of connecting certain nodes by an
  arrow. The purpose of these decorations is to encode the Cartan
  involution $\theta$. It can be reconstructed as follows. 

A black node stands for a simple root that is invariant under the
involution $\theta$ (and therefore, it will be a root of $G_c$). A
white node corresponds to a simple root that is not invariant under
$\theta$. White nodes can be connected by arrows; if the nodes
corresponding to $\alpha$ and $\beta$ are connected, this
signifies that $\alpha-\alpha^{\theta}=\beta-\beta^{\theta}$. As it
stands, this means that $\alpha$ and $\beta$ have the same component
on the subspace of eigenvalue $-1$ under $\theta$. Some reshuffling
teaches us that $\alpha-\beta$ belongs to the invariant
subspace. Hence a basis for the invariant subspace is formed by the
roots corresponding to black nodes, together with the differences of the
roots connected by arrows. On the orthogonal complement $\theta$ has
eigenvalue $-1$. This completes the specification of $\theta$, and
hence of the real form.

Note the two extremes: we can have diagrams of black dots only,
meaning $\theta=1$, and hence the real form is the compact one; a
diagram of white dots only, and no arrows indicates $\theta = -1$, and
hence the real form is the split form. 

\subsection{Group disintegration}

One might wonder if Satake diagrams have a role to play in ``group
disintegration'', the chain of subgroups of $G$ one finds in
oxidation. This is indeed the case.

\begin{figure}[ht]
\begin{center}
\includegraphics[width=8cm]{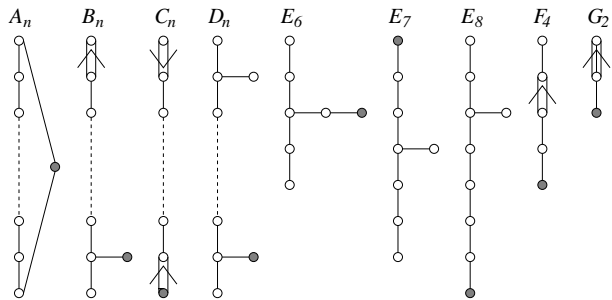}
\caption{Extended Dynkin diagrams; the marked node is the extended
  one. How to produce extended Satake diagrams from these is explained
  in the text.}\label{dynk} 
\end{center}
\end{figure}
Let us consider ``Extended Satake diagrams''; analogously to Satake
diagrams, we define these as Extended Dynkin diagrams with
decoration. We copy the decoration of all nodes from the Satake
diagram; only the decoration of the extended
node has to be specified. This is easy upon using the fact that the Satake
diagram specifies $\theta$ completely. The extended root is linearly
dependent on the other roots; the corresponding node \emph{can} only
be black if all the other nodes are black, and in that case, it
\emph{must} be black. There can be no arrows pointing to the
extended node; its partner would have to be contained in the Satake diagram,
contradicting that the Satake diagram completely encodes the
involution.

We know that the Extended Dynkin diagram encodes the regular subgroups
of $G$; as always we want to decompose in an $SL$ chain and a
complementary subgroup. It is an old observation \cite{Julia:1982gx,
  Cremmer:1999du} that group disintegration has to start at the end of
the Dynkin diagram where the affine vertex attaches; in
\cite{Keurentjes:2002xc} we argued that this is so because one can
find the U-duality group by taking the extended Dynkin diagram,
starting an $SL$ chain at the extended node, and erase the appropriate
nodes to disconnect it from the rest of the diagram.

Exactly the same recipe works for Extended Satake diagrams. Provided
we are not in the ``all black nodes'' case (corresponding to the
compact form, implying a trivial coset, and hence no oxidation), the
extended node is white. As $SL$ groups are
split, they are composed entirely of white nodes, hence we look for
chains of these. The nodes that we erase are not allowed to
be black; black nodes correspond to the invariant subspace, which
cannot become lower dimensional; the $SL$ roots are entirely in the
orthogonal complement. Lastly, the remaining diagram,
complementary to the $SL$ chain must make sense as a Satake
diagram. Sticking to these rules, the Satake diagram complementary to
the $SL$ chain gives the semi-simple part of the U-duality group. We
again have to complete by adding Abelian factors if the final (not
extended) diagram has less nodes than the (not extended) diagram we
started with. Erased nodes were always white, but there was the
possibility of some nodes being connected by arrows. A direct computation
reveals that a pair of erased nodes connected by an arrow gives a
$U(1)$-factor, while a surplus of erased white nodes without arrows
gives $\R$-factors. 

\begin{figure}[ht]
\begin{center}
\includegraphics[bb=100 400 500 820, width=6cm]{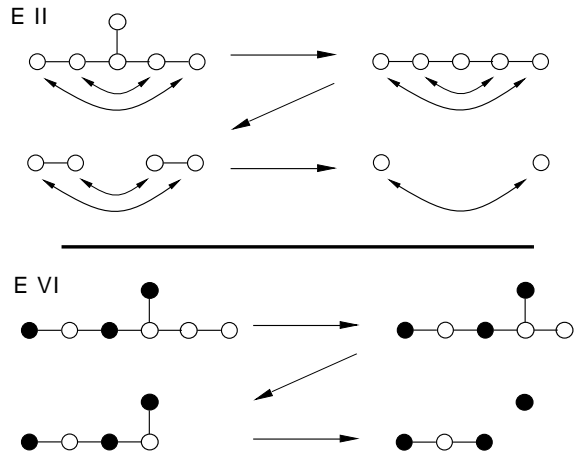}
\caption{Examples of group disintegration for Satake
  diagrams}\label{disfig} 
\end{center}
\end{figure}

We depicted 2 examples in figure \ref{disfig}. The first
example is $E_{6(2)}$, oxidizing in steps to 6 dimensions. The
extended node connects to the branch of the diagram not decorated by
arrows. The group disintegration proceeds along this branch leading to
the diagrams of $SU(3,3)$ and $SL(3,\C)$, which are the groups
following from an explict computation \cite{deWit:1992wf}. In the last
step we loose two nodes connected by an arrow; the remaining diagram
is $SL(2,\C) \cong SO(3,1)$, and as explained in the above, a
$U(1)$-factor should be added. 

The second oxidation chain depicts the
decompactification of the bosonic sector of $N=6$
supergravity. Subsequently we find the diagrams of $SO^*(12)$,
$SU^*(6)$ and $SU(2) \times SU^*(4)$. Note
that in the last step we find a disconnected black node, giving an
$SU(2)/SU(2)$ factor in the U-duality group. This 6 dimensional chiral
supergravity was considered in \cite{D'Auria:1997cz}. The equations of
motion and Bianchi identities appear not to have been written down,
but can be straightforwardly constructed from our formalism (for the
bosonic sector).

\section{Results for various cosets} \label{results}

We now apply the formalism of the previous section to sigma models
based on cosets of $G/H$, for two important classes of non-compact
$G$. The first class are simple Lie groups over the complex numbers,
the second consists of real forms of simple Lie groups. 
 
\subsection{Groups over the complex numbers}

Consider the complexification of a compact simple Lie algebra that
generates a group $H$. The complexified algebra  generates a group
that we will denote by $H^{\C}$. The maximal compact subgroup is the
compact form of $H$. The Satake diagrams for these consist of two
copies of the Dynkin diagram of $H$, with \emph{open} nodes, where the nodes
in the two copies are connected pairwise by arrows. Every symmetric
combination of generators of the two copies gives rise to a
non-compact generator, while every anti-symmetric combination results
in a compact generator. Therefore, the character $\sigma$ is always zero.

As is easily seen, the invariant subspace contains no roots, nor does
its complement. The group $G_s \times G_c = (\R \times U(1))^r$ (which
is a torus), with $r$ the rank of the algebra $H$ (and the $\R$-rank
of $H^{\C}$). Therefore, a sigma model in 3 dimensions
on $H^{\C}/H$ can never be oxidized. Note that it is certainly
possible for $H^{\C}$ to contain $SL$ subgroups, but these have to be
index 2 or higher, and therefore do not meet our constraints. A
well known example is $SL(2,\C)$, the double cover of the 4
dimensional Lorentz group $SO(3,1)$, which is known to be not
oxidizable.  
 
Next a systematic survey of all non-compact real forms of
simple Lie algebra's is presented. Many theories that we 
reconstruct from these are known as bosonic sectors from the context of
supergravity theories. The non-compact real forms have been classified
in chains, labelled by a capital and a roman number
\cite{Helgason}. The capital coincides with the one assigned to
the (complexified) algebra. For each non-compact real form we list its
Satake diagram, its maximal compact subgroup $H$, its character
$\sigma$, its $\R$-rank $r$, and the groups $G_s$, and $G_c$. In the
cases where the 3 dimensional sigma model is oxidizable,
we present the $U$-duality groups in the oxidation chain. To recover
the matter content of the theory, we refer the reader to the
decompositions in appendix B of \cite{Keurentjes:2002xc}; these can be
applied directly to the non-split non-compact forms, upon noting that
one only has to replace the split U-duality groups centralizing
$SL(D \! - \! 2,\R)$ by the non-compact forms in the tables in this
paper. Note that the oxidation chains for the split forms are always
longer than those of the non-split forms, so the table is
truncated for higher dimensions. 

For completeness we mention the split non-compact forms, but we will
omit most details about them, as these can be found in
\cite{Cremmer:1999du, Keurentjes:2002xc} and references therein. 
 
\subsection{$A_n$}

\begin{figure}[ht]
\begin{center}
\includegraphics[bb=0 340 500 760, width=8cm]{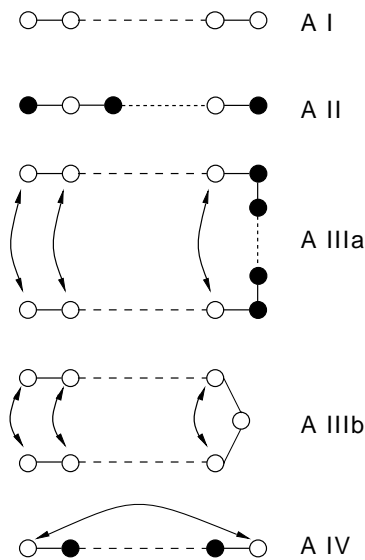}
\caption{Satake diagrams for non-compact forms of the $A_n$ groups}
\label{SA} 
\end{center}
\end{figure}

\subsubsection{$A\,I$: $SL(n\! + \!1,\R)$}

This is the split non-compact form the $A_n$ algebra. These oxidize to
general relativity in $n+3$ dimensions \cite{Cremmer:1997ct}, for the
$SL(2,\R)/SO(2)$ case this has been known for a long time. The maximal
compact subgroup, character, and $\R$-rank are 
$$
H = SO(n\! + \!1); \qquad \sigma = n; \qquad r=n.
$$ 
For split forms $G = G_s$ and $G_c$ is the trivial group:
$$
G_s \! \times \! G_c = SL(n\! + \!1,\R) \! \times \! \{ e \}
$$
For details we refer to \cite{Cremmer:1997ct}.

\subsubsection{$A\,I\!I$: $SU^*(n\! + \!1)$}

For this non-compact form we require $n > 1$ ($SU^*(2) =SU(2)$, the
compact form), and $n$ odd.  The maximal compact subgroup, character
and $\R$-rank are 
$$
H = Sp(\hlf(n\! + \!1)); \qquad \sigma = -n\! - \!2; \qquad r=\hlf(n\!
- \!1) 
$$
The subgroups $G_s$ and $G_c$ are found to be:
$$
G_s \! \times \! G_c = \R^{\hlf(n\! - \!1)} \! \times \!
SU(2)^{\hlf(n\! + \!1)} 
$$

As is obvious from $G_s$, sigma models based on these cosets cannot be
oxidized, due to the absence of a level 1 $SL(n)$-group. This was
already noted in \cite{Breitenlohner:1987dg}. 

\subsubsection{$A\,I\!I\!I$, $A\,I\!V$: $SU(n\! + \!1\! - \!p,p)$}

As obviously $SU(a,b)$ is isomorphic to $SU(a,b)$ we restrict to $p
\leq \hlf(n+1)$. In the list of Satake diagrams (fig \ref{SA}) we
have distinguished $A\,I\!I\!Ia$ with $1 < p < \hlf(n+1)$, and
$A\,I\!I\!Ib$ where $p=\hlf(n\! + \!1)$. The series $A\,I\!V$ refer to
the case $p=1$. Note that $SU(1,1) \cong SL(2,\R)$.  
 
The maximal compact subgroup, character and $\R$-rank are given by 
$$
H = S(U(n\! + \!1\! - \!p) \times U(p)); \qquad \sigma = 1-(n\! +
\!1\! - \!2p)^2; \qquad r=p 
$$

In computing $G_s$ and $G_c$ we have to distinguish between $A
\,I\!I\!Ia$ and $A\,\!I\!V$ on the one hand, and $A\,I\!I\!Ib$ on the
other hand. If $n\! + \!1 \neq 2p$ ($A \,I\!I\!Ia$ and $A\,\!I\!V$) 
$$
G_s \! \times \! G_c = SL(2,\R)^p \! \times \! SU(n\! + \!1\! - \!2p)
\! \times \! U(1)^p 
$$
while for  $n\! + \!1 = 2p$ ($A\,I\!I\!Ib$)
$$
G_s \! \times \! G_c = SL(2,\R)^p \! \times \! U(1)^{p-1}
$$

From $G_s$ we see that these theories always oxidize to
$4$ dimensions (with $SU(1,1) \cong SL(2,\R)$ fitting perfectly in
the pattern). The relevant groups for the oxidation are: 
$$
\ba{|r|r@{/}l|}
\hline
D & G & H \\
\hline
4 & SU(n\! - \!p,p\! - \!1) \! \times \! U(1) & 
S(U(n\! - \!p) \! \times \! U(p\! - \!1)) \! \times \! U(1)\\
3 & SU(n\! + \!1\! - \!p,p) &  S(U(n\! + \!1\! - \!p) \! \times \!
U(p)) \\ \hline
\ea
$$
The 4 dimensional theory contains general relativity, coupled to $n-1$
vectors, and a sigma model.  Note that the $U(1)$ factor cancels out
of the sigma model, though the vectors carry charges under the
$U(1)$. For $p=1$ the 4-d sigma model is empty. 

For $p=2$, and $SU(2,1)$ the symmetric spaces of the 3-d sigma model
are quaternionic and relevant in the context of $N=2$
supergravity. For these the process of oxidation is the inverse of the
$\mathbf{c}$-map \cite{deWit:1991nm}.  
 
\subsection{$B_n$} 

\subsubsection{$ B\,I$, $B\,I\!I$: $SO(2n\! + \!1\! -
  \!p,p)$}\label{Bsec} 

These models are very similar to those in based on $D\,I$, $D\,I\!I$
(which we will discuss in subsection \ref{Dsec}), and in the existing
literature are often treated simultaneously. As obviously $SO(a,b)
\cong SO(b,a)$, we restrict to $p \leq n$. In the list of Satake
diagrams we have distinguished between $B\,Ia$, with $p=n$ (the split
form), and $B\,Ib$ where $1 < p < n$. The case $B\,I\!I$ refers to
$p=1$.  

The maximal compact subgroup, character and $\R$-rank are given by
$$
H = SO(2n\! + \!1\! - \!p) \! \times \! SO(p); \qquad \sigma = n-2(n\!
- \!p)(n\! - \!p\! + \!1); \qquad r=p 
$$

For the split case $B\,Ia$ one has of course
$$
G_s \! \times \! G_c = SO(n+1,n) \! \times \! \{ e \},
$$
while for the other cases
$$
G_s \! \times \! G_c = SO(p,p) \! \times \! SO(2n\! - \!2p\! + \!1)
$$

The table of groups appearing in the oxidation chain is
\bd
\ba{|r|r@{/}l|}
\hline
D & G & H \\
\hline
p+2 &  SO(2n\! - \!2p\! + \!1) \! \times \! \R  & SO(2n\! - \!2p\! +
\!1) \\ 
\vdots & \multicolumn{2}{c|}{\vdots} \\
d+2 &  SO(2n\! - \!p\! - \!d\! + \!1,p\! - \!d) \! \times \! \R &
 SO(2n\! - \!p\! - \!d\! + \!1) \! \times \! SO(p\! - \!d)\\
\vdots & \multicolumn{2}{c|}{\vdots} \\
6 & SO(2n\! - \!p\! - \!3,p\! - \!4) \! \times \! \R & SO(2n\! - \!p\!
- \!3) \! \times \! SO(p\! - \!4)\\
5 & SO(2n\! - \!p\! - \!2,p\! - \!3) \! \times \! \R &  SO(2n\! -
\!p\! - \!2) \! \times \! SO(p\! - \!3) \\
4 & SO(2n\! - \!p\! - \!1,p\! - \!2) \! \times \! SL(2,\R) &  SO(2n\!
- \!p\! - \!1)  \! \times \! SO(p\! - \!2) \! \times \! SO(2) \\
3 & SO(2n\! - \!p\! + \!1,p) &  SO(2n\! - \!p\! +
\!1) \! \times \! SO(p) \\ \hline
\ea
\ed
Furthermore, for $p >3$ there is an alternative decomposition leading to
$$
\ba{|r|r@{/}l|}
\hline
D & G & H \\
\hline
6 & SO(2n\! - \!p\! - \!2,p\! - \!3) & SO(2n\! - \!p\! - \!2) \!
\times \! 
SO(p\! - \!3) \\
\hline
\ea
$$

An exceptional case is $n=p=1$, where one has $SO(2,1)\cong
SL(2,\R)$. For $SO(2q,1)$, $q\neq 1$, there is no oxidation possible. 

In the other cases the theory can be oxidized to $p\! + \!2$
dimensions, where one finds an Einstein-dilaton type gravity, coupled
to an antisymmetric tensor, and $2n\! - \!2p\! + \!1$ vectors. The
latter form the vector representation of $SO(2n\! - \!2p\! + \!1)$,
though this compact factor cancels from the sigma model.  

As for the split forms \cite{Keurentjes:2002xc} there is (for $p>3$) a
separate branch in $6$ dimensions, leading to a theory with general
relativity, a sigma model, and (anti-)self dual tensors. The number of
self dual and anti-self dual tensors is not equal, and the theory is
chiral, even in absence of fermions. 

As in the split case \cite{Keurentjes:2002xc} one can analyze the
explicit theory in $p+2$ dimensions, to show that the Bianchi identity
for the 2-form takes the form 
\be
\dif F_{(3)} = \hlf \sum_{i=1}^{2n-2p+1} F_{(2)i} \wedge F_{(2)i}
\ee
which is reminiscent of the identity $\dif H = \hlf \textrm{Tr}(F
\wedge F) - \hlf \textrm{Tr}(R \wedge R)$ from string theory. Indeed,
these models follows from the general considerations of
\cite{Narain:1985jj} for string theories. The models based on
$SO(8,2q+1)$ have supersymmetric extensions to theories with 16
supersymmetries; in 10 dimensions, these give type I supergravity
coupled to an odd number of Yang-Mills multiplets
\cite{Chamseddine:ez}. Though a priori of relevance to string theory,
there appear to be very few realizations of these models from superstrings
\cite{HJK}. 

Other interesting theories are the ones based on
$SO(4,2q+1)$ (note that $SO(4,1) \cong Sp(1,1)$) that give rise to
quaternionic manifolds in 3 dimensions, and are important in the
context of theories with $8$ supersymmetries \cite{deWit:1992wf}. The
2 different 6 dimensional branches correspond to exchanging vector
multiplets with tensor multiplets. A string theory realization can be
constructed by compactifying a heterotic string theory on $K3$, with a
number of pointlike instantons (5-branes) (and possibly
truncations). These give tensor multiplets for the $E_8 \times E_8$
string \cite{Ganor:1996mu}, and vector multiplets for the
$Spin(32)/\Z_2$ string \cite{Witten:1995gx}. Compactifying on an extra
circle, the 2 theories become T-dual, as reflected in our oxidation chain. 

\subsection{$C_n$}

\subsubsection{$C\,I$ :$Sp(n,\R)$}

This is the split real form, analyzed before in \cite{Cremmer:1999du}
and \cite{Keurentjes:2002xc}. 

Maximal compact subgroup, character and $\R$ rank are given by:
$$
H = U(n); \qquad \sigma = n; \qquad r=n
$$

As this is a split form, obviously
$$
G_s \! \times \! G_c = Sp(n,\R) \times \{ e \}
$$
These theories always oxidize to $4$ dimensions; see
\cite{Cremmer:1999du} for details. 

\subsubsection{$C\,I\!I$: $Sp(n\! - \!p,p)$}

As $Sp(a,b) \cong Sp(b,a)$, we restrict to $p \leq \hlf n$.  In the
list of Satake diagrams, we have furthermore distinguished between
$C\,I\!Ia$, for $p \neq \hlf n$, and $C\,I\!Ib$ for $p = \hlf n$. 
 
Maximal compact subgroup, character and $\R$-rank are given by
$$
H = Sp(p) \! \times \! Sp(n\! - \!p); \qquad \sigma = -n-2(n - 2p)^2
\qquad r=p 
$$

In all cases $G_s \times G_c$ can be computed to be 
$$
G_s \! \times \! G_c = (SL(2,\R)_2 \times SU(2)_2)^p \! \times \!
Sp(n\! - \!2p) 
$$
We have added the subscript $2$ to the $SL(2,\R)_2$ and $SU(2)_2$
subgroups, to indicate that these are index 2 subgroups; they have
short roots. There is no $SL(n,\R)$ subgroup with index 1, i.e. with
long roots, and hence these theories do not oxidize. 

The 3-d cosets with $p=1$ are quaternionic spaces \cite{deWit:1991nm}. 

\subsection{$D_n$}

\subsubsection{$D\,I$, $D\,I\!I$: $SO(2n\! - \!p,p)$} \label{Dsec}

These models are similar to those based on $B\,I$, $B\,I\!I$
(subsection \ref{Bsec}), and are often treated simultaneously in the
literature. 

Again, obviously $SO(a,b) \cong SO(b,a)$, and we restrict to $p \leq
n$. The list of Satake diagrams distinguishes between $D\,Ia$, where
$p=n$ (the split form), $D\,Ib$ where $p = n-1$, and $D\,Ic$, with $1
< p < n-1$. Finally $D\,I\!I$ refers to $p=1$. 

The maximal compact subgroup, character and $\R$-rank are given by
$$
H = SO(2n\! - \!p) \! \times \! SO(p); \qquad \sigma = n-2(n\! -
\!p)^2; \qquad r=p 
$$

In all cases we have
$$
G_s \! \times \! G_c = SO(p,p) \! \times \! SO(2n\! - \!2p)
$$

The analysis of these theories is very similar to the one for the $B
\, I$ and $B \,I\!I$ theories. The table of groups encountered in
group disintegration is 
\bd
\ba{|r|r@{/}l|}
\hline
D & G & H \\
\hline
p+2 & SO(2n\! - \!2p) \! \times \! \R  & SO(2n\! - \!2p) \\
 \vdots & \multicolumn{2}{c|}{\vdots} \\
d+2 & SO(2n\! - \!p\! - \!d,p\! - \!d) \! \times \! \R & SO(2n\! -
\!p\! - \!d) \! \times \! SO(p\! - \!d)\\
\vdots & \multicolumn{2}{c|}{\vdots} \\
6 & SO(2n\! - \!p\! - \!4,p\! - \!4) \! \times \! \R &   SO(2n\! -
\!p\! - \!4) \! \times \! SO(p\! - \!4)\\
5 & SO(2n\! - \!p\! - \!3,p\! - \!3) \! \times \! \R & SO(2n\! - \!p\!
- \!3)  \! \times \!  SO(p\! - \!3) \\
4 & SO(2n\! - \!p\! - \!2,p\! - \!2) \! \times \! SL(2,\R) & SO(2n\! -
\!p\! - \!2)  \! \times \!  SO(p\! - \!2)\! \times \! SO(2) \\
3 & SO(2n\! - \!p,p) &  SO(2n\! - \!p) \! \times \! SO(p) \\
\hline
\ea
\ed
Furthermore, there is an extra possibility if $p >3$:
$$
\ba{|r|r@{/}l|}
\hline
D & G & H \\
\hline
6 & SO(2n\! - \!p\! - \!3,p\! - \!3) & SO(2n\! - \!p\! - \!3)\! \times
\! SO(p\! - \!3)  \\
\hline
\ea
$$
Finally, for $n=p=1$ we have $SO(1,1)\cong \R$, which does not lead to
an oxidizable theory. For $SO(2q+1,1)$, oxidation is
never possible (note that $SO(3,1) \cong SL(2,\C)$ and $SO(5,1) \cong
SU^*(4)$). 

In the other cases the theory can be oxidized to $p\! + \!2$
dimensions, where one finds an Einstein-dilaton type gravity, coupled
to an antisymmetric tensor, and $2n\! - \!2p$ vectors. The latter form
the vector representation of $SO(2n\! - \!2p)$, though this compact
factor cancels from the sigma model.  

There is (for $p>3$) a separate branch in $6$ dimensions, leading
to a theory with general relativity, a sigma model, and (anti-)self
dual tensors. The number of self dual and anti-self dual tensors is
only equal for $p=n$, hence for split forms. For the non-split real
forms, the 6 dimensional theory is chiral. 

As in the $B\,I$ and $B\,I\!I$ cases, in $p+2$ dimensions the Bianchi
identity for the 2-form takes the form 
\be
\dif F_{(3)} = \hlf \sum_{i=1}^{2n-2p} F_{(2)i} \wedge F_{(2)i}
\ee
again to be compared with the identity $\dif H = \hlf \textrm{Tr}(F
\wedge F) - \hlf \textrm{Tr}(R \wedge R)$ in string theory (but do
note the possibility that $p=n$). Also these models follow from the
general considerations of \cite{Narain:1985jj} for string
theories. For theories with 16 supersymmetries the models based on
$SO(8,2q)$ are important, giving type I supergravity
 coupled to an even number of Yang-Mills multiplets in 10
dimensions \cite{Chamseddine:ez}. 

The theories based on $SO(4,2q)$ (note that
$SO(4,2) \cong SU(2,2)$) give rise to quaternionic manifolds in 3
dimensions, and are important in the context of theories with $8$
supersymmetries \cite{deWit:1992wf}.   As in the $B$ chains, a string
theory realization can be constructed by compactifying a heterotic
string theory on $K3$, with pointlike instantons (5-branes). These give tensor
multiplets for the $E_8 \times E_8$ string \cite{Ganor:1996mu}, and
vector multiplets for  the $Spin(32)/\Z_2$ string
\cite{Witten:1995gx}. These theories are T-dual after compactification
on an extra circle, as reflected in our oxidation chain.

\subsubsection{$D\,I\!I\!I$: $SO^*(2n)$}

This non-compact form has maximal compact subgroup and character 
$$
H = U(n); \qquad \sigma = -n.
$$

There is an important difference between $n$ even or odd.
If $n$ is even, the Satake diagram is of type $D\,I\!I\!Ia$ and the
$\R$-rank $r=n/2$. In this case, the groups $G_s \times G_c$ are given by: 
$$
G_s \! \times \! G_c = SL(2)^{n/2} \! \times \! SU(2)^{n/2}.
$$
If $n$ is odd, the Satake diagram is of type $D\,I\!I\!Ib$ and the
$\R$-rank $r=(n-1)/2$. Then the groups $G_s \times G_c$ are given by 
$$
G_s \! \times \! G_c = SL(2)^{\hlf(n-1)} \! \times \!
SU(2)^{\hlf(n-1)} \! \times \! U(1) 
$$

Irrespective of whether $n$ is even or odd, the table of groups
appearing in oxidation is: 
$$ 
\ba{|r|r@{/}l|}
\hline
D & G & H \\
\hline
4 & SO^*(2n\! - \!4) \! \times \! SU(2) & U(n\! - \!2) \! \times \!
SU(2)\\ 
3 & SO^*(2n) & U(n)\\
\hline
\ea
$$
The maximal dimension is always 4. The 4 dimensional theory has a
graviton, and a sigma model on $SO^*(2n-4)/U(n-2)$. There are $2n-4$
vectors, organizing with their duals in vectors of $SO^*(2n-4)$, and a
doublet of the (hidden) $SU(2)$ factor. A special case is $SO^*(4)
\cong SL(2,\R) \times SU(2)$, but the previous statements still hold
for these theories. 

\subsection{$E_6$}

\subsubsection{$E\,I$: $E_{6(6)}$}

This is the split form. Its maximal compact subgroup, character and
$\R$-rank are: 
$$
H = Sp(4); \qquad \sigma = 6 \qquad r=6.
$$

As obvious for split form, the group $G_c$ is trivial:
$$
G_s \! \times \! G_c = E_{6(6)} \times \{ e \}
$$
The analysis of \cite{Cremmer:1999du} (see also
\cite{Keurentjes:2002xc}) revealed that this theory can be
oxidized to 8 dimensions.

\subsubsection{$E\,I\!I$: $E_{6(2)}$}

This real form has maximal compact subgroup, character and $\R$-rank
given by 
$$
H = SU(6) \! \times \! SU(2); \qquad \sigma = 2; \qquad r=4.
$$
The groups $G_s$ and $G_c$ are given by
$$
G_s \! \times \! G_c = SO(4,4)\times U(1)^2
$$
The chain of groups appearing in the oxidation is
$$
\ba{|r|r@{/}l|}
\hline
D & G & H \\
\hline
6 & SL(2,\C) \times U(1) & SU(2) \times U(1) \\
5 & SL(3,\C) &  SU(3)\\
4 & SU(3,3) &  S(U(3) \! \times \! U(3)) \\
3 & E_{6(2)} &  SU(6) \! \times \! SU(2) \\
\hline
\ea
$$
The 6 dimensional theory consists of general relativity, a sigma model
on $SL(2,\C)/SU(2)$, 4 vectors, and 2 2-tensors. Actually this theory
is a close relative of the 6 dimensional theory that can be oxidized from
$F_{4(4)}/(Sp(3)\times SU(2))$ \cite{Cremmer:1999du}. The group $E_6$
allows an outer automorphism, that is manifest in its Dynkin
diagram. Decorating the Dynkin diagram with its outer automorphism,
one precisely finds the Satake diagram $E\,I\!I$. Quotienting $E_6$ by
its outer automorphism, one obtains $F_{4}$, and a computation reveals
that the real form of $F_{4}$ embedded in $E_{6(2)}$ is $F_{4(4)}$. As
pointed out in \cite{Cremmer:1999du}, the 6 dimensional theory
oxidized from $F_{4(4)}$ is an extension of a class of theories
studied by Sagnotti \cite{Sagnotti:1992qw}. We expect that the present 6
dimensional theory, oxidized from $E_{6(2)}$, allows a similar
treatment.  

The lower dimensional theories in the oxidation chain
can also be found in \cite{deWit:1992wf}.  

\subsubsection{$E\,I\!I\!I$: $E_{6(-14)}$}

This non-compact form has maximal compact subgroup, character and $\R$-rank  
$$
H = SO(10) \! \times \! U(1); \qquad \sigma = -14; \qquad r=2
$$
The groups $G_s$ and $G_c$ can be computed to be
$$
G_s \! \times \! G_c = SL(2)^2 \! \times \! SU(4)
$$
It oxidizes to 4 dimensions. The relevant groups are given by
$$
\ba{|r|r@{/}l|}
\hline
D & G & H \\
\hline
4 & SU(5,1) &  U(5) \\
3 & E_{6(-14)} & SO(10) \! \times \! U(1) \\
\hline
\ea
$$
This theory allows a supersymmetric extension: it corresponds to the
bosonic sector of $N=5$ supergravity. The 4 dimensional model has a
graviton, a sigma model on $SU(5,1)/U(5)$, and 10 vectors, that,
together with their duals, form the 20 dimensional antisymmetric
3-tensor irrep of $SU(5,1)$. 

\subsubsection{$E\,I \!V$: $E_{6(-26)}$}

This non-compact form has maximal compact subgroup, character and $\R$-rank
given by 
$$
H=F_4; \qquad \sigma = -26; \qquad r =2.
$$
Computation of the groups $G_s$ and $G_c$ leads to
$$
G_s \! \times \! G_c = \R^2 \! \times \! SO(8)
$$
As $G_s$ has no $SL(n,\R)$ subgroup, this theory cannot be oxidized.

\subsection{$E_7$}

\subsubsection{$E \, V$: $E_{7(7)}$}

This is the split form, that was analyzed in
\cite{Cremmer:1999du}. Maximal compact subgroup, character and
$\R$-rank are given by 
$$
H = SU(8); \qquad \sigma = 7; \qquad r=7
$$
The groups $G_s$ and $G_c$ are 
$$
G_s \! \times \! G_c = E_{7(7)} \times \{ e \}
$$
as obvious for a split form

This theory oxidizes to $10$ dimensions, but has an additional branch
in $8$ dimensions. For details we refer the reader to
\cite{Cremmer:1999du, Keurentjes:2002xc}.  

\subsubsection{$E\,V\!I$: $E_{7(-5)}$}

This non-compact real form has maximal compact subgroup, character and
$\R$-rank given by
$$
H = SO(12) \! \times \! SU(2); \qquad \sigma = -5; \qquad r=4
$$
The groups $G_s$ and $G_c$ are given by
$$
G_s \! \times \! G_c = SO(4,4) \! \times \! SU(2)^3
$$
We immediately see that this theory oxidizes to 6 dimensions. In the
oxidation chain one finds the $U$-duality groups:
$$
\ba{|r|r@{/}l|}
\hline
D & G & H \\
\hline
6 & SU(2) \! \times \! SU^*(4) & SU(2) \! \times \! Sp(2) \\
5 & SU^*(6) & Sp(3)\\
4 & SO^*(12) & U(6)\\
3 & E_{7(-5)} &  SO(12) \! \times \! SU(2) \\  
\hline
\ea
$$ 
In 6 dimensions the theory oxidizes to the bosonic sector of a
chiral $(2,1)$ supergravity, as conjectured in
\cite{Julia:1980gr, Julia:1982gx} and demonstrated in
\cite{D'Auria:1997cz}. The amount of
supersymmetry in 6 dimensions is $(2,1)$, and the chiral nature of the
theory complicates its analysis. Its defining equations
appear to have not been written down. With our present formalism it is
straighforward to reconstruct the equations for the bosonic sector of
the 6-d theory.

The same bosonic sector can be built from an $N=2$ theory (lower
dimensional theories in the oxidation chain were discussed in
\cite{deWit:1992wf}). This theory is closely related to the $E_{6(2)}$
and $F_{4(4)}$ theories; a projection of the $E_{7(-5)}$ root lattice 
on the space $\cH_-$ defined by the Cartan involution gives the root
lattice of $F_{4(4)}$, e.g. all these 3 theories have the same
restricted root system with different multiplicities for the roots
\cite{Helgason}.   

\subsubsection{$E\,V\!I\!I$: $E_{7(-25)}$}

This real form has maximal compact subgroup, character and $\R$-rank
given by
$$
H = E_6 \! \times \! U(1); \qquad \sigma = -25; \qquad r=3.
$$
A computation reveals that $G_s$ and $G_c$ are given by
$$
G_s \! \times \! G_c = SL(2)^3 \! \times \! SO(8)
$$
We immediately see that the theory can be oxidized to $4$
dimensions. The relevant U-duality groups are
$$
\ba{|r|r@{/}l|}
\hline
D & G & H \\
\hline
4 & SO(10,2) & SO(10) \! \times \! SO(2)\\
3 &  E_{7(-25)} & E_6 \! \times \! U(1) \\
\hline  
\ea
$$
In 4 dimensions, we find general relativity, coupled to 20 scalars in
the coset, and 16 vectors that together with their duals transform in
the 32 dimensional spinor irrep of $SO(10,2)$.

\subsection{$E_8$}

\begin{figure}[ht]
\begin{center}
\includegraphics[bb=0 680 500 800, width=8cm]{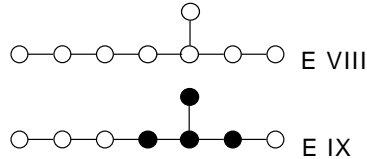}
\caption{Satake diagrams for non-compact forms of the $E_8$ groups}
\end{center}
\end{figure}

\subsubsection{$E\,V\!I\!I\!I$: $E_{8(8)}$}

This famous coset oxidizes to the bosonic sector of 11 dimensional
supergravity \cite{Cremmer:1978km}, and has a seperate branch in 10
dimensions, where it corresponds to the bosonic sector of IIB gravity
\cite{Schwarz:qr}. 

Its maximal compact subgroup, character and $\R$-rank,
are given by
$$
H = SO(16); \qquad \sigma=8; \qquad r=8
$$
As the group is split, we obviously have 
$$
G_s \! \times \! G_c = E_{8(8)} \times \{ e \}
$$
The literature on theories in this chain is immense. Important
original references are \cite{Cremmer:1978km, Schwarz:qr,
  Cremmer:1979up, Marcus:1983hb}, while a review is \cite{Obers:1998fb}.

\subsubsection{$E\,I\!X$: $E_{8(-24)}$}

The algebra $E_8$ has a second non-compact form, with maximal compact
subgroup, character and $\R$-rank given by
$$
H = E_7 \! \times \! SU(2);\qquad \sigma = 6; \qquad r=6
$$
The groups $G_s \times G_c$ are given by
$$
G_s \! \times \! G_c = SO(4,4) \! \times \! SO(8)
$$

This theory oxidizes to 6 dimensions, and the U-duality groups are
given by:
$$
\ba{|r|r@{/}l|}
\hline
D & G & H \\
\hline
6 &  SO(9,1) & SO(9)  \\
5 &  E_{6(-26)} &  F_4 \\
4 &  E_{7(-25)} &  E_6 \! \times \! U(1) \\
3 &  E_{8(-24)} &  E_7 \! \times \! SU(2) \\
\hline 
\ea
$$
In 6 dimensions, this theory includes general relativity, 9 scalars on
$SO(9,1)/SO(9)$, 16 vectors transforming as a spinor of $SO(9,1)$, and
5 2-tensors. 

The theories in this chain are well known in the context of the $\mathbf{r}$
and $\mathbf{c}$ maps \cite{deWit:1992wf}. The group $E_{8(-24)}$ is
again closely related to the groups $E_{7(-5)}$, $E_{6(2)}$ and
$F_{4(4)}$; all have the restricted root system (the roots projected on
the invariant subspace $\cH_-$) of $F_{4(4)}$, with different
multiplicities for the roots. All allow supersymmetric extensions.

Another amusing observation is that the
3 dimensional theories based on cosets formed by dividing $F_{4(4)}$,
$E_{6(2)}$, $E_{7(-5)}$ and $E_{8(-24)}$ by their maximal
compact subgroups, all oxidize to 6 dimensions, with sigma models on
$SL(2,\R) \cong SO(2,1)$, $SL(2,\C) \cong SO(3,1)$, $SL(2,\h) \cong
SO(5,1)$ and $SL(2, \mathbb{O}) \cong SO(9,1)$ respectively (see
e.g. \cite{Baez:2001dm} for definitions of $SL(2,\h)$ and $SL(2,
\mathbb{O})$). These are all Lorentz groups; in the supersymmetric
extensions of these theories this is dictated by $D=6$
(1,0) supersymmetry. The 2-tensors plus their duals transform in the
vector representation of the Lorentz groups (also dictated by
supersymmetry), while the vectors transform in a spinor
representation. A lot of mathematical structure in the
\emph{bosonic} sector of these theories is familiar from
supersymmetric Yang-Mills theories. Together with the links to
division algebra's and exceptional groups (see also
\cite{Gunaydin:1983rk}), these oxidation chains
provide interesting and entertaining mathematics. 

\subsection{$F_4$}

\subsubsection{$F\,I$: $F_{4(4)}$}

This split form has maximal compact subgroup, character and $\R$-rank
$$
H = Sp(3) \! \times \! SU(2); \qquad \sigma = 4; \qquad r=4;
$$
and
$$
G_s \! \times \! G_c = F_{4(4)} \times \{e\}.
$$
It oxidizes to 6 dimensions. An extensive discussion of the 6
dimensional theory is found in \cite{Cremmer:1999du}. Lower
dimensional theories from this chain can be found in
\cite{deWit:1992wf}. 

\subsubsection{$F\,I\!I$: $F_{4(-20)}$}

This non-compact form has maximal compact subgroup, character and
$\R$-rank 
$$
H = SO(9); \qquad \sigma = -20; \qquad r=1
$$
The groups $G_s \times G_c$ are computed to be
$$
G_s \! \times \! G_c = SL(2, \R)_2 \! \times \! Sp(3).
$$
The subscript 2 on $SL(2,\R)$ denotes that we are dealing with an
index 2 subgroup here. From this we immediately deduce that this
theory cannot be oxidized, as was already known since
\cite{Breitenlohner:1987dg}.

\subsection{$G_2$}

\subsubsection{$G$: $G_{2(2)}$}

The only non-compact form of the group $G_2$ is the split form, with
maximal compact subgroup, character and $\R$-rank
$$
H=SO(4); \qquad \sigma =2; \qquad r=2
$$
The groups $G_s$ and $G_c$ are obviously given by
$$
G_s \! \times \! G_c = G_{2(2)} \! \times \! \{ e \}
$$
This theory oxidizes to the bosonic sector of simple supergravity in 5
dimensions. For an extensive study of the theories in
this chain see \cite{Mizoguchi:1998wv}.

\section{Summary and remarks}

We have extended the analysis of \cite{Keurentjes:2002xc} to cover
oxidation from all coset theories formulated on $G/H$, with $G$ a
non-compact simple Lie group, and $H$ its maximal compact subgroup.

For these we had to deal with the Cartan involution. In the previous
paper we had been able to ignore the Cartan involution, because for
\emph{all} split groups the Cartan involution can be chosen such that
it acts as $\mathbf{-1}$ on the Cartan sub-algebra. As demonstrated in
this paper, dealing with generic Cartan involutions hardly poses any
problems; using some technology from group theory, we hardly need
additional ingredients for the sigma model analysis.

An important ingredient in our analysis were the compact subgroups
$G_c$. In particular, these were vital to our analysis of coset sigma
models, that forms the basis of our oxidation recipe. They can be
computed from the Satake diagram of the non-compact real form. Satake
diagrams are also helpful in reading of the process of group
disintegration; this basically follows the same pattern as for the
split groups, where the analysis is done in terms of Dynkin
diagrams. The extra decoration that accompanies the Satake diagram
encodes the non-compact forms we find in the various dimensions

Together with the analysis in \cite{Keurentjes:2002xc}, the results of
the present paper represent an exhaustive analysis. To some extent,
because many of the theories discussed here were known before, our
main contribution is the demonstration that we have exhausted all
possibilities for oxidation from simple Lie groups. We remind the
reader that this does however rely on the assertion that the higher
dimensional theories follow from the possible index 1 $SL(D \! - \! 2,\R)$
subgroups \cite{Keurentjes:2002xc}. We have explained that this is
equivalent to demanding a theory with exactly 1 graviton, and other
irreps allowing interpretations as form fields, and scalars. As there
exist no-go theorems on theories with multiple gravitons, and massless
fields that are not forms, this seems a reasonable requirement.  

An elegant result is that the full bosonic sector of a
large class of theories, among which many supergravity theories is
encoded in a surprisingly small set of ingredients:
essentially a Satake diagram (which compactly encodes the algebra, the
roots, the Cartan involution, the relevant subalgebra's and possible
physical dualities) and a set of equations: the dilaton equation
(\ref{cartaneqmo2}), a single equation relating the form fields and axions
(\ref{oxieq}), algebraic equations (\ref{alg}) for some axions and
dilatons, and the Einstein equation, which states that the Einstein
tensor couples to the energy momentum tensors of a set matter fields
(which matter fields follows again from group theory). 

A perhaps interesting observation is that in the formalism we used,
the theories based on non-split groups can always be recovered from
the split cases. We start with the split group, and build the theories
with the recipe of \cite{Keurentjes:2002xc}. Because of the choice of
(positive root) gauge, a theory based on a non-split form can
immediately be recovered by setting some fields to zero (see equation
(\ref{alg})). This is possible because the truncation of the model based
on the split group is equivalent to fixing the gauge in the other
model; essentially we are turning some non-compact generators into
compact ones, and then gauge them away. As it is the group $G_c$
determining the set of fields in question, this once more emphasizes
the important role this subgroup plays. 

The wide range of applicability of the methods developed in
\cite{Keurentjes:2002xc} and this paper leads us to expect that they
might be useful to other problems in the context of general
relativity and supergravity. We hope to report on other applications
in the future.

{\bf Acknowledgements:}  A project with A.~Hanany and B.~Julia \cite{HJK}
inspired me to research the issues presented here and in
\cite{Keurentjes:2002xc}. I thank Bernard Julia, Niels Obers, Pierre
Henry-Labordere and Louis Paulot for discussions, and Marc Henneaux
for lending a copy of \cite{Helgason}. The author is
supported in part by the ``FWO-Vlaanderen'' through project G.0034.02,
in part by the Federal office for Scientific, Technical and Cultural
Affairs trough the Interuniversity Attraction Pole P5/27 and in part
by the European Commision RTN programme HPRN-CT-2000-00131, in which
the author is associated to the University of Leuven.

\end{document}